\documentclass[prc,aps,float,nofootinbib,superscriptaddress,twocolumn]{revtex4-1}
\usepackage{amssymb}
\usepackage{amsmath}
\usepackage{amsfonts}
\usepackage{epsfig}
\usepackage{bbm}
\usepackage{comment}

\usepackage{multirow}
\usepackage{longtable}
\usepackage{array}
\usepackage{subfigure}
\usepackage{graphics}

\begin{document}

\title{Combining symmetry breaking and restoration with configuration interaction: \\ a highly accurate many-body scheme applied to the pairing Hamiltonian}
  
\author{J. Ripoche} \email{julien.ripoche@cea.fr}
\affiliation{Institut de Physique Nucl\'eaire, IN2P3-CNRS, Universit\'e Paris-Sud, Universit\'e Paris-Saclay, F-91406 Orsay Cedex, France}
\affiliation{CEA, DAM, DIF, F-91297 Arpajon, France}

\author{D. Lacroix} \email{lacroix@ipno.in2p3.fr}
\affiliation{Institut de Physique Nucl\'eaire, IN2P3-CNRS, Universit\'e Paris-Sud, Universit\'e Paris-Saclay, F-91406 Orsay Cedex, France}

\author{D. Gambacurta}\email{danilo.gambacurta@eli-np.ro}
\affiliation{Extreme Light Infrastructure - Nuclear Physics (ELI-NP),  M\u agurele, Romania}

\author{J.-P. Ebran} \email{jean-paul.ebran@cea.fr}
\affiliation{CEA, DAM, DIF, F-91297 Arpajon, France}

\author{T. Duguet} \email{thomas.duguet@cea.fr} 
\affiliation{KU Leuven, Instituut voor Kern- en Stralingsfysica, 3001 Leuven, Belgium}
\affiliation{CEA, DRF/IRFU/SPhN, Universit\'e Paris-Saclay, 91191 Gif-sur-Yvette, France}
\affiliation{National Superconducting Cyclotron Laboratory and Department of Physics and Astronomy, Michigan State University, East Lansing, MI 48824, USA}

%Finland}
\date{\today}

\begin{abstract} 
\begin{description}
\item[Background] Ab initio many-body methods have been developed over the past ten years to address mid-mass nuclei. In their best current level of implementation, their accuracy is of the order of a few per cent error on the ground-state correlation energy. Recently implemented variants of these methods are operating a breakthrough in the description of medium-mass {\it open-shell} nuclei at a polynomial computational cost while putting state-of-the-art models of inter-nucleon interactions to the test.
\item[Purpose]  As progress in the design of inter-nucleon interactions is made, and as questions one wishes to answer are refined in connection with increasingly available experimental data, further efforts must be made to tailor many-body methods that can reach an even higher precision for an even larger number of observable/quantum states/nuclei. It is the objective of the present work to contribute to such a quest by designing and testing a new many-body scheme.
\item[Methods] We formulate a truncated configuration interaction method that consists of diagonalizing the Hamiltonian in a highly truncated subspace of the total $N$-body Hilbert space. The reduced Hilbert space is generated via the particle-number projected BCS state along with projected seniority-zero two and four quasi-particle excitations. Furthermore, the extent by which the underlying BCS  state breaks $U(1)$ symmetry is optimized {\it in presence} of the projected two and four quasi-particle excitations. This constitutes an extension of the so-called restricted variation after projection method in use within the frame of multi-reference energy density functional calculations. The quality of the newly designed method is tested against exact solutions of the so-called attractive pairing Hamiltonian problem.
\item[Results] By construction, the method reproduce exact results for $N=2$ and $N=4$. For $N=(8,16,20)$, the error on the ground-state {\it correlation energy} is less than (0.006, 0.1, 0.15) \% across the entire range of inter-nucleon coupling defining the pairing Hamiltonian and driving the normal-to-superfluid quantum phase transition. The presently proposed method offers the advantage to automatically access the low-lying spectroscopy, which it does with high accuracy.
\item[Conclusions] The numerical cost of the newly designed variational method is polynomial ($N^6$) in system size. It achieves an unprecedented accuracy on the ground-state correlation energy, effective pairing gap and one-body entropy as well as on the excitation energy of low-lying states of the attractive pairing Hamiltonian. This constitutes a strong enough motivation to envision its application to realistic nuclear Hamiltonians in view of providing a complementary, accurate and versatile ab initio description of mid-mass open-shell nuclei in the future. 
\end{description}
\end{abstract}



\maketitle

\section{Introduction}
\label{Intro}

Methods to solve the $N$-body Schroedinger equation must cope with two specific attributes of inter-nucleon interactions that are responsible for the non-perturbative character of the nuclear many-body problem~\cite{Bogner:2006tw,Ramanan:2007bb}. The first trait relates to the strong inter-nucleon repulsion at short distances that translates into large off-diagonal coupling between states characterized by low and high (relative) momenta, i.e. the first element of non-perturbative physics is of ultra-violet nature and manifests itself in all nuclei independently of the detail of their structure. The second trait relates to the unnaturally large nucleon-nucleon scattering length in the S-wave/spin-singlet channel and with the tendency of inter-nucleon interactions to induce strong angular correlations between nucleons in the internal frame of the nucleus. This second element of non-perturbative physics is of infra-red character and only manifests itself in sub-categories of nuclei, i.e. in singly open-shell and doubly open-shell nuclei.

Off-diagonal coupling between low and high (relative) momenta can be tamed-down, at the price of inducing (hopefully weak) higher-body forces, by pre-processing the nuclear Hamiltonian via, e.g., a unitary similarity transformation~\cite{Bogner:2009bt}. Based on the transformed Hamiltonian, dynamical correlations\footnote{The denomination of {\it dynamical} and {\it non-dynamical} correlations is presently used in the quantum chemistry sense.} can be dealt with at a polynomial cost via standard many-body techniques typically based on systematic particle-hole-type expansions. Corresponding ab-initio methods, i.e. many-body perturbation theory (MBPT)~\cite{shavitt09a}, coupled cluster (CC) theory~\cite{Hagen:2013nca}, self-consistent Green's function (SCGF) theory~\cite{Dickhoff:2004xx,Cipollone:2013zma}, in-medium similarity renormalization group (IMSRG) theory~\cite{Hergert:2015awm}, have been developed and implemented with great success in the last ten years to deal with doubly-(sub)closed shell nuclei and their immediate neighbors.

Strong, i.e. non-dynamical, correlations induced in singly and doubly open-shell nuclei are of different nature and require specific attention. Several routes are possible, including full configuration interaction (CI) techniques~\cite{Jansen:2014qxa,Bogner:2014baa}. To proceed on the basis of methods whose cost scales polynomial with the number interacting nucleons, one option consists in exploiting the spontaneous breaking of symmetries induced by non-dynamical correlations at the mean-field level. This rationale allows one to incorporate a large part of the non-perturbative physics into a single product state that can serve as a reference for many-body expansions dealing efficiently with dynamical correlations. While traditionally developed within the frame of effective nuclear mean-field (i.e. energy density functional) approaches~\cite{bender03b,yannouleas07a,Duguet:2013dga}, this idea has been recently embraced to develop and implement ab initio Gorkov SCGF~\cite{soma11a,Soma:2013vca,Lapoux16}, multi-reference IMSGR~\cite{Hergert:2013uja,Hergert:2016etg} and Bogoliubov CC~\cite{Signoracci:2014dia} many-body techniques to tackle pairing correlations\footnote{While the formation of cooper pairs is primarily driven by the unnaturally large nucleon-nucleon scattering length in the spin-singlet isospin-triplet channel, it is also partly due to indirect processes associated with the exchange of collective vibrations~\cite{barranco04a,Idini:2011zz,Lesinski:2008cd,duguet13BCS}.}. This is achieved by allowing the reference state to break $U(1)$ global gauge symmetry associated with particle-number conservation. While the restoration of the broken symmetry, eventually necessary in any finite quantum system, has been formulated for MBPT~\cite{LaCroixGambacurta,Duguet16PNRBCC} and CC techniques~\cite{Duguet16PNRBCC}, it has only been implemented so far in the context of nuclear ab initio calculations via MR-IMSGR techniques~\cite{Hergert:2013uja,Hergert:2016etg}. 

Methods based on a symmetry breaking reference state are currently allowing a breakthrough in the ab initio description of (singly) open-shell nuclei and are putting state-of-the-art inter-nucleon interactions to the test~\cite{Soma:2013xha,Hergert:2014iaa,Lapoux16}. In the most advanced truncation schemes implemented so far, this is achieved by allowing a few percent error on the ground-state correlation energy\footnote{We are only quoting here the systematic uncertainty associated with the truncation of the many-body expansion scheme.}. As progress on inter-nucleon interactions is made, and as the questions one wishes to answer are refined in connection with increasingly available experimental data, further efforts must be made to tailor many-body methods (with minimized numerical costs) that can reach higher precision along with more observable/quantum states/nuclei. It is the objective of the present work to design and test a new many-body scheme that has the potential to do so. 

In order to characterize the potential of new many-body schemes, one can test them against solutions of exactly solvable many-body Hamiltonians. To be in position to draw meaningful conclusions, the schematic Hamiltonian must be significantly non-trivial and capture enough key physics of the real system of interest. In view of the above discussion, we presently focus on the so-called attractive pairing Hamiltonian~\cite{Ric64,Ric66,Ric68,Duk04} whose main merit is to effectively model the superfluid character of finite nuclear systems or any other mesoscopic fermionic superfluid system. More specifically, the dynamic of $N$ interacting fermions is governed by the Hamiltonian
\begin{eqnarray}
H(g) & \equiv & \sum_{k=1}^{\Omega} e_k (a^\dagger_k a_k + a^\dagger_{\bar k} a_{\bar k} )
- g  \sum_{k \neq l}^\Omega  a^\dagger_k a^\dagger_{\bar k} a_{\bar l}  a_l  \, , \label{eq:tb} 
\end{eqnarray}
where $\Omega$ denotes the number of doubly degenerate ($e_k=e_{\bar k}$) time-reversed\footnote{The conjugation of the two single-particle states can actually originate from any dichotomic symmetry such as time reversal, signature or simplex.} single-particle states $(k,{\bar k})$ characterized by the creation operators $(a^\dagger_k, a^\dagger_{\bar k})$. The double degeneracy of single-particle states is meant to mimic (even-even) doubly open-shell nuclei treated via the spontaneous breaking of $SO(3)$ rotational symmetry, i.e. exploiting explicitly the concept of deformation. In the present study, the distance between successive pairs of degenerate levels is constant, i.e. $e_{k+1}-e_k \equiv \Delta e$, and the system is systematically studied at "half-filling", i.e. $N=\Omega$. Modeling, e.g., rare-earth nuclei, one typically has $\Delta e\sim 500$\,keV. The coupling strength $g \in [0,+\infty[$ characterizes the attractive pairing interaction that scatters pairs of nucleons from any given set of degenerate single-particle states to any other set with a constant probability amplitude. As $g$ increases, the system is known to undergo a phase transition from a normal to a superfluid system  at a critical value $g=g_c$ that depends on the number of particles $N$. Eventually, the relevant parameter of the model is the ratio $g/\Delta e$ that measures the pairing strength relative to the spacing between successive pairs of single-particle states. For rare-earth nuclei, one typically\footnote{Throughout the paper, numerical values quoted for $g$ are in unit of $\Delta e$, i.e. they actually corresponds to quoting the ratio $g/\Delta e$.} has $g/\Delta e\sim 0.5$.

While the eigenstates of this Hamiltonian can be obtained exactly via direct diagonalization, i.e. full CI~\cite{Vol01,Zel03,Vol07}, Quantum-Monte Carlo simulations~\cite{Cap98,Muk11} or the numerical solution of so-called Richardson equations~\cite{Ric64,Ric66,Ric68,Duk04,Van06,claeys15a}, there exists a long history of search for accurate approximate solutions at the lowest possible algorithmic cost. Indeed, the numerical cost of exact methods scales factorial with $N$ or $\Omega$, which quickly becomes prohibitive for realistic systems of interest. Among these approximate methods\footnote{We only focus here on methods that can be applied systematically for all coupling strength $g$, i.e. before, across and after the normal-to-superfluid phase transition. If not, more calculations could be mentioned, including those based on the self-consistent random phase approximation~\cite{Duk03} applicable to $g<g_c$.} are the variation after particle-number projection Bardeen-Cooper-Schrieffer approach (VAP-BCS) approach~\cite{Die64,deG66,egido82a,egido82b,blaizot86,heenen93a,sheikh00,anguiano01b,San08,Hup11}, truncated CI calculations~\cite{Mol97,Mol07,Pil05}, CC calculations without~\cite{Duk03,Jon13,Lim13,Ste14,Henderson:2014vka-b} or with $U(1)$ symmetry breaking~\cite{Henderson:2014vka,Signoracci:2014dia}.  

Particular attention must be paid to the highly accurate method recently proposed in Refs.~\cite{Duk16,Deg16}. Reconciling the performance of CC doubles in the normal phase with the merit of VAP-BCS in the strongly interacting regime, this method, coined as polynomial similarity transformation (PoST), reaches less than 1 \% error
\begin{eqnarray}
(\Delta E / E)_c & = & \left( 1 - \frac{E^{\rm approx}_c}{E^{\rm exact}_c}\right)\times 100 ~({\rm in} ~ \%)
\end{eqnarray}   
on the ground-state correlation energy $E_c$ defined as the total energy minus the Hartree-Fock (HF) energy obtained by filling the $N$ lowest levels, for all interaction strength and moderate particle number~\cite{Deg16}. 

In view of this recent development, we presently wish to design a many-body scheme that scales polynomial with the number of interacting fermions and whose results display an error on the ground-state correlation energy that is better than 1$\%$ for any interaction strength. To reach this ambitious objective, the variational approach introduced below builds on Ref.~\cite{LaCroixGambacurta} and combines two key characteristics
\begin{enumerate}
\item $U(1)$ symmetry breaking and restoration
\begin{enumerate}
\item spontaneous
\item optimized
\end{enumerate}
\item truncated CI diagonalization.
\end{enumerate}
While Ref.~\cite{LaCroixGambacurta} displayed encouraging results by exploiting spontaneous $U(1)$ symmetry breaking and restoration within a perturbative approach, the present work strongly improves on them by exploiting truncated CI techniques and by optimizing the extent by which the symmetry is broken prior to being restored. In addition, a strong asset of the presently proposed method is to provide a highly accurate account of low-lying excited states. The approach being based on a wave-function ansatz, observable besides the energy can easily be accessed as exemplified by the computation of the effective pairing gap and the one-body entropy. 

The paper is organized as follows. Section~\ref{Formalism} displays the formalism in such a way that several standard methods can be easily recovered as particular cases. Sections~\ref{spontaneous}-\ref{Observables} provide extensive numerical results and compare them to exact solutions as well as to those obtained from existing approximate methods. Eventually, results for low-lying excited states are discussed. Section~\ref{Conclu} concludes the present work and elaborates on some of its perspectives.

\section{Formalism}
\label{Formalism}

\subsection{Basis construction}

We first consider the BCS solution for $H(g)$\footnote{It is implicitly assumed here that the Hamiltonian is replaced by the grand potential $H(g) -\lambda A$, with $\lambda$ the chemical potential  used to impose that the BCS solution has the right number   of particles in average. The particle number operator is  $A=\sum_{k=1}^{\Omega} (a^\dagger_k a_k + a^\dagger_{\bar k} a_{\bar k} )$.} carrying even number-parity as a quantum number. It can be written as
\begin{equation}
\label{e:bogvac}
| \Phi (g) \rangle \equiv \prod_{k=1}^{\Omega} \left( u_k (g) + v_k(g) \, a^\dagger_k a^\dagger_{\bar  k}\right)| 0 \rangle  \, ,
\end{equation}
where the coefficients $(u_k(g), v_k(g))$, satisfying $u^2_k(g)+ v^2_k(g)=1$ for all $k$, are obtained by solving standard BCS equations~\cite{ring80a}. Quasi-particle creation operators, whose hermitian conjugates annihilate $| \Phi (g) \rangle $, are obtained via the BCS transformation
\begin{subequations}
\label{BCStransformation}
\begin{eqnarray}
\beta^\dagger_k(g) &\equiv& u_k(g) \, a^\dagger_k - v_k(g) \, a_{\bar k} \, , \\
\beta^\dagger_{\bar k}(g) &\equiv& u_k(g) \, a^\dagger_{\bar k}  + v_k(g) \, a_{k} .
\end{eqnarray}   
\end{subequations}
Normal-ordering $H(g)$ with respect to $| \Phi (g) \rangle $ allows one to rewrite it under the form
\begin{equation}
\label{split1}
H(g) = H_{0}(g) + H_{1}(g) \, ,
\end{equation} 
where the unperturbed part reads as
\begin{equation}
H_{0}(g) = {\cal E}_0 (g) + \sum_{k=1}^{\Omega} E_k(g) \left( \beta^\dagger_k \beta_k +
  \beta^\dagger_{\bar k} \beta_{\bar k} \right) \label{eq:hqp} .
\end{equation} 
The real number ${\cal E}_0 (g)$ denotes the approximate BCS ground-state energy whereas 
\begin{equation}
E_k(g) \equiv \sqrt{(e_k-\lambda)^2+\Delta^{2}(g)}\, ,
\end{equation} 
defines BCS quasi-particle energies, with $\Delta(g)$ the BCS pairing gap~\cite{Bri05}. The explicit expression of the residual interaction $H_{1}(g)$ can be obtained accordingly~\cite{ring80a}.

The BCS vacuum and the set of quasi-particle (qp) excitations built on top of it
\begin{eqnarray}
| \Phi^{kl\ldots}(g) \rangle &\equiv& \beta^{\dagger}_{k}(g) \, \beta^{\dagger}_{l}(g) \,  \ldots   |  \Phi(g) \rangle  \, , 
\end{eqnarray}
form a complete eigenbasis ${\cal  B}(g)$ of $H_{0}(g)$ over Fock space ${\cal  F}$ such that
\begin{eqnarray*}
H_{0}(g)\, |  \Phi(g) \rangle &=& {\cal E}_0(g) \, |  \Phi(g) \rangle \, , \\
H_{0}(g) \, |  \Phi^{kl\ldots}(g) \rangle &=& \left[{\cal E}_0 (g) \!+\! E_{k}(g) \!+\! E_{l}(g)\!+\!\ldots\right] |  \Phi^{kl\ldots}(g) \rangle  \label{phi} \, . 
\end{eqnarray*}
Being interested in eigenstates of even-even systems with seniority zero, the only basis states of actual interest are those involving pairs of time-reversed quasi-particle excitations for which the shorthand notation 
\begin{equation}
| \Phi^{k\bar{k}l\bar{l}\ldots}(g) \rangle \equiv | \Phi^{\breve{k}\breve{l}\ldots}(g) \rangle
\end{equation}
is used. Eventually, all basis states can be written as BCS vacua of the form
\begin{eqnarray}
| \Phi^{\breve{k}\breve{l}\ldots}(g) \rangle &=& \prod_{m=1}^{\Omega} \left( u^{\breve{k}\breve{l}\ldots}_m (g) + v^{\breve{k}\breve{l}\ldots}_m(g) \, a^\dagger_m a^\dagger_{\bar  m}\right)| 0 \rangle \, .
\end{eqnarray}
This notation implicitly includes the BCS vacuum as a particular case when using the BCS coefficients $(u_m(g),v_m(g))$. For the excited state $| \Phi^{\breve{k}\breve{l}\ldots}(g) \rangle$, one has 
$u^{\breve{k}\breve{l}\ldots}_m(g) \equiv u_m(g)$ and $v^{\breve{k}\breve{l}\ldots}_m(g) \equiv v_m(g)$, except for $m=k,l,\ldots$ for which $u^{\breve{k}\breve{l}\ldots}_m(g) \equiv -v_m(g)$ 
and $v^{\breve{k}\breve{l}\ldots}_m(g) \equiv u_m(g)$.

While the eigenstates of $H_{0}(g)$ form a complete orthonormal basis of Fock space, they break $U(1)$ symmetry associated with particle number conservation, i.e. they are not eigenstates of the particle number operator $A$. In order to recover states belonging to the Hilbert space ${\cal  H}_N$ associated with the physical number $N$ of nucleons in the system, a projection operator
\begin{equation}
P_N =  \frac{1}{2\pi}\int_{0}^{2\pi} \! d{\varphi} \; \,e^{i\varphi (A-N)} \, , \label{projector}
\end{equation} 
can be applied to generate the set of projected qp excitations
\begin{equation}
| \Phi^{\breve{k}\breve{l}\ldots}_N(g) \rangle \equiv P_N  | \Phi^{\breve{k}\breve{l}\ldots}(g) \rangle \, \label{projstates}
\end{equation} 
forming a non-orthogonal overcomplete basis ${\cal  B}_N(g)$ of ${\cal  H}_N$. While  $| \Phi^{\breve{k}\breve{l}\ldots}_N(g) \rangle$ directly originates from $| \Phi^{\breve{k}\breve{l}\ldots}(g) \rangle$, it is worth noting that the former is not an eigenstate of $H_0(g)$.

For $g>g_c$, each basis state defined through Eqs.~\ref{e:bogvac}-\ref{projstates} builds in the breaking of the $U(1)$ symmetry prior to performing its exact restoration. As such, each state $| \Phi^{\breve{k}\breve{l}\ldots}_N(g) \rangle$ is a complex entanglement of 0p-0h, 2p-2h, $\cdots$, Np-Nh excitations with respect to the HF reference state, as nicely illustrated by Eq.~(5) of Ref.~\cite{Duk16}. For $g<g_c$, however, the BCS vacuum actually reduces to the HF reference state such that each state  $| \Phi^{\breve{k}\breve{l}\ldots}_N(g) \rangle$ identifies with one n-particle/n-hole (np-nh) excitation on top of it belonging to ${\cal H}_N$\footnote{In this case, the action of $P_N$ is superfluous such that this is already true of the unprojected basis states $| \Phi^{\breve{k}\breve{l}\ldots}(g) \rangle$.}. It is worth noting that certain combinations of qp excitations do not actually have any np-nh counterpart in ${\cal H}_N$ for $g<g_c$. For example, a 2qp excitation of time-reversed states tend towards a Slater determinant belonging to  ${\cal H}_{N\pm2}$ below $g_c$.

\subsection{Truncated CI method}
\label{truncCI}

We wish to approximate eigenstates of $H(g)$, starting with its ground state, via an exact diagonalization within the subpace of ${\cal  H}_N$ spanned by a subset of states of ${\cal  B}_N(g)$. A similar idea was used in a different context on the basis of projected QRPA states~\cite{Gam12} to estimate transfer probabilities between many-body states with different particle numbers. In the present case, eigenstates are approximated by the  ansatz\footnote{The coefficient of the projected BCS vacuum is not set a priori because we keep the possibility to remove it altogether from the variational ansatz, in which case $c=0$.}
\begin{eqnarray}
| \Psi_N (g) \rangle & \equiv & c \, | \Phi_N(g) \rangle \nonumber \\
&& + \sum_{k} c_{\breve{k}} \, | \Phi^{\breve{k}}_N(g) \rangle \nonumber \\
&& + \sum_{l<m} c_{\breve{l}\breve{m}} \, | \Phi^{\breve{k}\breve{m}}_N(g) \rangle \, , \label{eq:var1}
\end{eqnarray}
i.e. it mixes the particle-number-projected BCS vacuum with projected 2qp and 4qp excitations. The number of states in the linear combination is
\begin{eqnarray}
n_{\rm st} & =& n_{0{\rm qp}} + n_{2{\rm qp}} + n_{4{\rm qp}}   \nonumber \\
&=& 1 + C_{\Omega}^{1}   + C_{\Omega}^{2}   \nonumber \\
&=& 1 + \Omega + \frac{\Omega (\Omega -1)}{2}  \, , \label{eq:number}
\end{eqnarray}
with $N=\Omega$ in the present application.

The many-body state is determined variationally 
\begin{eqnarray}
\delta \left\{ \langle \Psi_N (g) |H(g)| \Psi_N (g) \rangle - E(g) \langle \Psi_N (g) | \Psi_N (g) \rangle  \right\} & = & 0 \, , \nonumber
\end{eqnarray}
where the minimization is performed with respect to the set of coefficients $\{c^*_{\alpha}\} \equiv \{c^*, c^*_{\breve{k}}, c^*_{\breve{l}\breve{m}} \}$, where $\alpha$ scans all states in the linear combination defining $| \Psi_N (g) \rangle$ in Eq.~\ref{eq:var1}. This leads to $n_{\rm st}$ coupled equations\footnote{Given that $P_N$ is a projector ($P^2_N=P_N$) and that $H(g)$ commutes with it ($[P_N,H(g)]=0$), it is sufficient to apply the projector on only one of the two states involved in any matrix element of the overlap or Hamiltonian matrices. This is the reason why we omitted the subscript $N$ in the bra $\langle \Phi^\alpha (g) |$ entering Eq.~\ref{diago1}.}
\begin{equation}
\sum_\beta c_\beta \langle \Phi^\alpha (g) | H(g) | \Phi_N^\beta (g) \rangle =  E(g) \sum_\beta c_\beta \langle \Phi^\alpha (g) | \Phi_N^\beta (g) \rangle \, . \label{diago1} 
\end{equation}
Matrix elements of $H(g)$ between the basis states as well as the overlap between the latter can be estimated using standard projection techniques. Explicit forms are given in the appendix of Ref.~\cite{LaCroixGambacurta}. 

Equation~\ref{diago1} is nothing but the Schroedinger equation represented in a finite-size non-orthogonal basis. It is solved by first diagonalizing the overlap matrix through a unitary transformation ${\cal U}_N(g)$
\begin{equation}
\sum_\beta \langle \Phi^\alpha (g) | \Phi_N^\beta (g) \rangle \, {\cal U}^{\beta\zeta}_N(g) = n^{\zeta}_N(g) \, {\cal U}^{\alpha\zeta}_N(g) \, , \label{diago2} 
\end{equation}
leading to a new set of orthonormal states
\begin{equation}
| \Theta_N^\zeta (g) \rangle \equiv  \sum_{\alpha} \frac{{\cal U}^{\alpha\zeta}_N(g)}{n^{\zeta}_N(g)}| \Phi_N^\alpha (g) \rangle \, , \label{diago3} 
\end{equation}
that is eventually used to diagonalize $H(g)$. The number of new orthonormal states is of course equal to $n_{\rm st}$. However, the size of the basis must actually be reduced prior to diagonalizing $H(g)$ by removing states with eigenvalues below a chosen threshold $\epsilon$, i.e. states that encode the redundancy of the initial non-orthogonal overcomplete basis. We will illustrate this point in Sec.~\ref{redundancy}.

\subsection{Particular cases}
\label{subcases}

One must note that the above scheme incorporates several existing approaches as particular cases
\begin{enumerate}
\item When limiting ansatz~\ref{eq:var1} to the sole first term, one recovers the particle-number projection after variation BCS (PAV-BCS) method. In this case, there is obviously no diagonalization to perform.
\item For $g<g_c$, i.e. in the normal phase, the scheme reduces to a standard truncated CI method~\cite{Mol97,Mol07,Pil05}, limited to 2p-2h configurations in the present case\footnote{As mentioned above, 2qp excitations of time-reversed states have no counterpart in ${\cal H}_N$ below $g_c$. Consequently, corresponding coefficients $c_{\breve{k}}$ are identically zero by construction in such a case.}. In this case, the number of states does not comply with Eq.~\ref{eq:number}, i.e. it is replaced by 
\begin{eqnarray}
n_{\rm st} & =& n_{0p0h} + n_{2p2h}   \nonumber \\
&=& 1 + \left(C_{\Omega/2}^{1}\right)^2   \nonumber \\
&=& 1+ \frac{\Omega^2}{4}  \, , \label{eq:number}
\end{eqnarray}
which for large $\Omega$, corresponds to essentially half of the cardinal defined in Eq.~\ref{eq:number}. The space spanned by the truncated basis is thus not continuous  through $g_c$. Consequences will be discussed in Sec.~\ref{spontaneous}.
\item When computing the mixing coefficients $\{c_{\alpha}\}$ from second-order (particle-number {\it unprojected}) MBPT, the diagonalization step is avoided. The residual interaction $H_{1}(g)$ contains terms with four quasi-particle operators\footnote{Terms with two creation or two annihilation operators are zero when $|  \Phi(g) \rangle$ satisfies the BCS equations~\cite{Signoracci:2014dia}, i.e. in Moller-Plesset MBPT~\cite{MPMBPT}.}~\cite{ring80a}. Consequently, $H_1(g)$ only couples the BCS vacuum $| \Phi(g) \rangle$ to 4qp excitations $| \Phi^{\breve{k}\breve{l}}(g) \rangle$ at second order. As a result, coefficients $c_{\breve{k}}$ associated with 2qp excitations are identically zero at that order. Ansatz~\ref{eq:var1} can be both implemented in {\it absence} of particle-number projection, in which case one works within a standard MBPT scheme, or in {\it presence} of the particle-number projection, in which case one works within a particle-number projected MBPT scheme that we can coin as MBPT$_N$\footnote{The mixing coefficients $\{c_{\alpha}\}$ are still computed from MBPT without particle number projection. Note that an alternative particle-number-restored MBPT based on a projective formula has been recently proposed~\cite{Duguet16PNRBCC} but not yet applied.}. Of course, standard second-order MBPT based on a HF reference state is recovered from MBPT$_N$ at $g<g_c$. It happens that MBPT and MBPT$_N$ have been applied to the pairing Hamiltonian in Ref.~\cite{LaCroixGambacurta} and serve as an inspiration for the generalizations introduced in the present work. Corresponding results will be briefly reminded in Sec.~\ref{spontaneous}.
\end{enumerate}

\subsection{Optimized order parameter}
\label{optorderparam}

Let us introduce one additional level of improvement. At a given value of the coupling strength $g$, the states forming the non-orthogonal overcomplete basis ${\cal  B}_N(g)$ have been naturally built so far from the BCS solution $| \Phi (g) \rangle$ of $H(g)$. Consequently, the extent by which $| \Phi (g) \rangle$ (possibly) break $U(1)$ symmetry, as characterized by its pairing gap $\Delta(g)$, is in one-to-one correspondence with the coupling $g$ defining the physical Hamiltonian. However, it is not at all obvious that the subpart of the resulting basis ${\cal  B}_N(g)$ used in the truncated CI calculation optimally captures the physics of the Hamiltonian $H(g)$. 

At each "physical" value $g$, it is thus possible to foresee the diagonalization of the Hamiltonian $H(g)$ in the (0qp, 2qp, 4qp) subpart of ${\cal  B}_N(g_{{\rm aux}})$ associated with an {\it auxiliary} value $g_{{\rm aux}}$, i.e. with the basis built from the BCS solution $| \Phi (g_{{\rm aux}}) \rangle$ of an auxiliary pairing Hamiltonian $H(g_{{\rm aux}})$\footnote{To some extent, performing standard truncated CI calculations based on a basis of np-nh Slater determinants already exploits this idea when dealing with $H(g)$ with $g>g_c$, i.e. it is nothing but using the basis ${\cal  B}_N(g_{{\rm aux}})$ built from the reference state corresponding to $g_{{\rm aux}}<g_c$ in connection with a  Hamiltonian $H(g)$ defined by $g>g_c$.}. Following this line of thinking, one can scan all values $g_{{\rm aux}} \in [0,+\infty[$ in order to find the optimal auxiliary coupling $g_{{\rm opt}}$. This extra step consists of spanning a larger manifold of states than when working at $g_{{\rm aux}}=g$. The method is thus of variational character, i.e. the optimal auxiliary coupling $g_{{\rm opt}}$ is obtained at the minimum of the curve $E^{g_{{\rm aux}}}(g)$ produced by repeatedly applying the truncated CI calculation, i.e. by solving Eq.~\ref{diago1} for the Hamiltonian $H(g)$ while varying the auxiliary coupling $g_{{\rm aux}}$ defining the basis states.

This scheme extends the so-called restricted variation-after-projection (RVAP) method designed within the frame of symmetry-restored nuclear energy density functional calculations~\cite{Rod05}. Generically speaking, the idea is to scan the symmetry-restored energy as a function of a collective variable that monitors the extent by which the unprojected reference state breaks the symmetry. In the present case of $U(1)$ symmetry, this order parameter is nothing but the pairing gap $\Delta(g_{{\rm aux}})$ associated with the BCS reference state $| \Phi (g_{{\rm aux}}) \rangle$. Typically, tuning the value of the gap can be done by solving BCS equations while adding a Lagrange constrain term. In the present case, however, $\Delta(g_{{\rm aux}})$ is a monotonic function of $g_{{\rm aux}}$ (see Fig.~\ref{fig:effgap}) such that one can directly use $g_{{\rm aux}}$ as a collective variable and solve for $H(g_{{\rm aux}})$. 

The novelty of the presently proposed scheme is that the optimal order parameter $g_{{\rm opt}}$ of the reference state is not only determined in presence of the symmetry restoration but also in presence of the mixing with projected 2pq and 4qp states, i.e. at the level the truncated CI calculation itself. As discussed below, this significantly impact the value of  $g_{{\rm opt}}$ and the associated quality of the variational ansatz.

\section{Truncated CI calculations}
\label{spontaneous}

\subsection{Perturbation theory}
\label{mbpt}

\begin{figure}[htbp] 
\includegraphics[width=\linewidth]{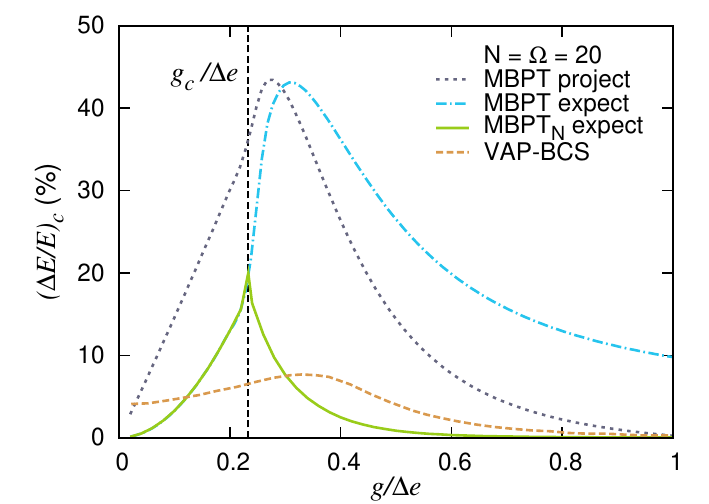} 
\caption{(color online) Error $(\Delta E / E)_c$ on the ground-state correlation energy as a function of $g$ for $N=\Omega=20$. Results are shown for (i) second-order particle-number unprojected MBPT energy based on a projective formula (grey dotted line), (ii) second-order particle-number unprojected MBPT energy based on a hermitian expectation value formula (blue dashed-dotted line), (iii) second-order particle-number projected MBPT$_N$ energy based on a hermitian expectation value formula (green solid line) and (iv) for VAP-BCS~\cite{Duk16} (brown dashed line). The critical value $g_c$ is indicated by the black dashed vertical line.}
\label{fig1:juju} 
\end{figure}

For reference, we first illustrate MBPT and MBPT$_N$ methods employed in Ref.~\cite{LaCroixGambacurta} and briefly introduced in Sec.~\ref{subcases} above. Second-order results are displayed in Fig.~\ref{fig1:juju} for $N=20$ and compared with VAP-BCS results~\cite{Duk16}. Three main lessons can be learnt from these calculations
\begin{enumerate}
\item Second-order corrections avoid systematically the collapse of the correlation energy that occurs as $g$ decreases through $g_c$ in BCS or PAV-BCS calculations~\cite{San08}. Of course, the VAP-BCS method also avoids the collapse but at the price of a significantly more sophisticated calculation.
\item Particle-number projection drastically improves over unprojected results. Given that standard Rayleigh-Schroedinger MBPT is based on a projective energy formula while MBPT$_N$ of Ref.~\cite{LaCroixGambacurta} relies on a hermitian expectation value, we also display the latter in absence of the projection in order to disentangle its effect. For $g<g_c$, the improvement solely comes from using the expectation value formula given that the reference state does not break $U(1)$ symmetry in the first place and thus the symmetry restoration cannot have any effect. For $g>g_c$, one sees that using the expectation value formula does not improve the results by itself and even deteriorates MBPT results obtained from a projective formula. Thus, the very significant improvement seen in the solid line does originate from the particle-number projection.
\item Except in the vicinity of $g_c$, MBPT$_N$ results are better than VAP-BCS, both in weak and strong coupling regimes. In particular, results display the correct limit as $g \rightarrow 0$ contrary to VAP-BCS calculations~\cite{Duk16}. It is not surprising given that standard second-order MBPT theory is known to converge to the right limit as $g$ tends to zero. More surprisingly, MBPT$_N$ converges very rapidly towards exact results as $g$ increases beyond $g_c$. As a matter of fact, for $g > 0.6$, results are even better than the very accurate PoST approach of Ref.~\cite{Deg16} that, by construction, matches VAP-BCS in the strong pairing regime.
\end{enumerate}

The quality of these results obtained at a low computational cost over both weakly- and strongly-coupled regimes teaches us that the space spanned by the states involved, i.e. the particle-number projected BCS vacuum and  particle-number projected 4qp excitations, contain key information to treat the physics of superfluid systems. Indeed, the spontaneous breaking of the symmetry, followed by its further restoration, allows one to resum non-dynamical correlations efficiently  whereas corrections associated with 4qp excitations seem to capture a large part of the dynamical correlations. Still, results are significantly above the 1 \% error on the correlation energy that constitutes our present objective. 

One natural generalization of the approach would be to include higher-order perturbative corrections. However, the rapid increase of the dimensionality of the probed Hilbert space translates in a severe augmentation of the computational cost. Alternatively, we move from a perturbative to a non-perturbative approach  via a diagonalization method while keeping the dimensional of the probed Hilbert space essentially the same.  

\subsection{Diagonalization}
\label{diag}

At each $g$, $H(g)$ is diagonalized within the space spanned by the non-orthogonal set of projected 0qp, 2qp and 4qp states built out of the BCS state $| \Phi (g) \rangle$, as explained in Sec.~\ref{truncCI}. The calculation reduces, as discussed in Sec.~\ref{subcases}, to a diagonalization in a truncated basis made of 0p-0h and 2p-2h configurations built out of the HF reference state for $g<g_c$. 

\begin{figure}[htbp] 
\includegraphics[width=\linewidth]{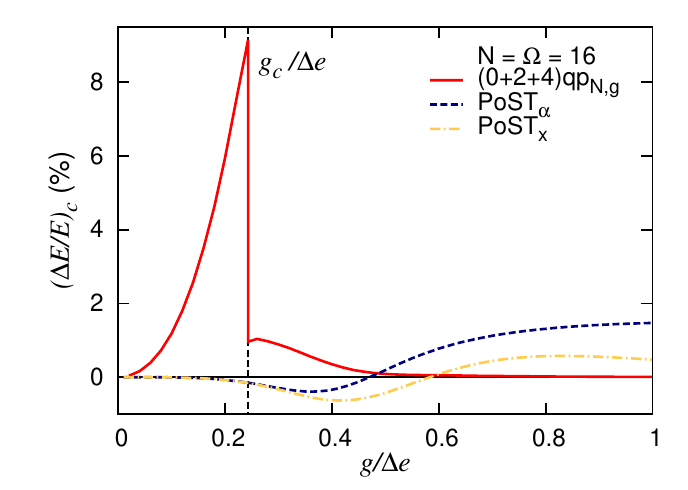} 
\caption{(color online) $(\Delta E / E)_c$ as a function of $g$  for $N=\Omega=16$. Results are shown for the truncated CI (red solid line),  PoST$_\alpha$~ \cite{Deg16} (dark blue dashed line) and PoST$_x$~ \cite{Deg16} (yellow dashed-dotted line) calculations.}
\label{fig2:juju} 
\end{figure}  

The error on the correlation energy is displayed in Fig.~\ref{fig2:juju} for $N=\Omega=16$. The diagonalization greatly improves the accuracy for $g>g_c$ compared to the perturbative calculation discussed above. The error is below the targeted 1 \% for all coupling beyond $g_c$ and quickly drops far below it as $g$ moves away from the BCS threshold. Contrarily, results from the truncated CI calculation are similar to the perturbative calculation below the threshold. Eventually, a discontinuity of the result occurs at $g=g_c$. 

The last feature can be qualitatively understood from the discontinuity of the basis dimension as $g_c$ is approached from below or from above, as already alluded to in Sec.~\ref{subcases}. By construction, the basis contains 0p-0h and 2p-2h Slater determinants of ${\cal H}_N$ below $g_c$. While a subset of 4qp states converges towards the 2p-2h Slater determinants when approaching $g_c$ from above, others become more and more dominated by Slater determinants belonging to Hilbert spaces associated with neighboring (even) number of particles. Still, residual components corresponding to np-nh Slater determinants of  ${\cal H}_N$ are extracted from them by projection. Consequently, the limit of the truncated CI calculation as $g_c$ is approached from above corresponds to a standard truncated CI calculation associated with a basis containing higher-order np-nh Slater determinants beyond 2p-2h configurations, the basis size being approximately twice the one below threshold. This feature greatly improves the quality of the method and illustrates the benefit of starting from symmetry broken (and restored) basis states above $g_c$. 

Eventually the proposed method is competitive with the PoST method of Ref.~\cite{Deg16} and becomes even quickly superior as one enters the strongly coupled regime. Still, the strict reduction of the method to a truncated CI based on sole 0p-0h and 2p-2h configurations below $g_c$ is not sufficient to reach the desired accuracy across both  normal and  superfluid phases and to obtain a smooth description throughout the transition. In Sec.~\ref{Optimized} below, this intrinsic limitation is overcome  while further improving the accuracy for all $g$. Before discussing this additional level of improvement, let us  first focus on the redundant character of the basis and of the optimal set of qp configurations one should start from.

\subsection{Basis redundancy and qp configurations}
\label{redundancy}

Based on unprojected MBPT, it is natural to first add 4qp excitations to the BCS reference state in the variational ansatz $| \Psi_N (g) \rangle$. The argument that the BCS reference state is not coupled to 2qp excitations via the residual  interaction $H_1(g)$ does not however stand once the particle-number projector is inserted; thus, our additional inclusion of 2qp excitations in the variational state $| \Psi_N (g) \rangle$. The final set of projected 0qp, 2qp and 4qp states is not orthonormal and thus contains a certain degree of redundancy. As explained in Sec.~\ref{truncCI}, this requires the diagonalization of the overlap matrix to extract a subset $n_{\rm sub} \leq n_{\rm st}$ of relevant orthonormal states characterized by large enough eigenvalues $n^{\zeta}_N(g) \geq \epsilon$. Let us now typify the relevant states depending on the original set of configurations included in the variational ansatz and characterize at the same time the quality of the associated results. We still focus on the $N=16$ case. 

\begin{figure}[htbp] 
\includegraphics[width=\linewidth]{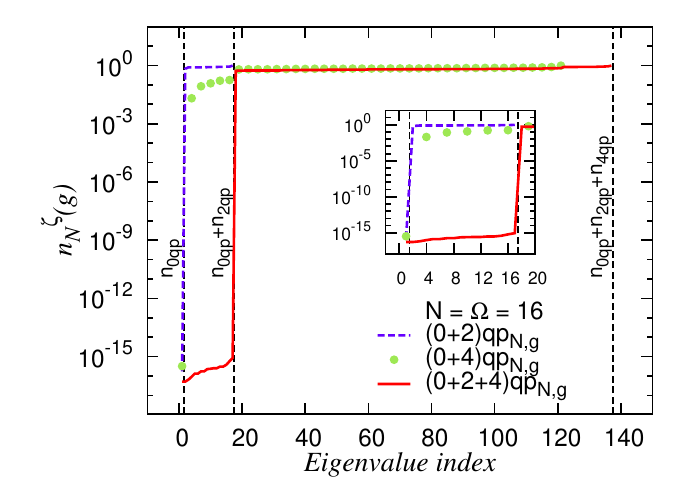} 
\caption{(color online) Eigenvalues $n^{\zeta}_N(g)$ (relative to the largest of them) of the overlap matrix (cf. Eq.~\ref{diago2}) ordered in increasing values for $N=16$ and $g/\Delta e = 0.8$. A logarithmic scale is used for the vertical axis. Results for $| \Psi_N (g) \rangle$  made of projected 0qp, 2qp and 4qp (red solid line) configurations, made of projected 0qp and 4qp configurations (green filled circles) or made of projected 0qp and 2qp configurations (purple dashed line) are shown.} % in the main plot. Results for $| \Psi_N (g) \rangle$ made of projected 0qp and 2qp configurations are also shown in the insert.}
\label{eigenvaluesnorm} 
\end{figure}

The upper panel of Fig.~\ref{eigenvaluesnorm} displays the $n_{\rm st} = 1 + 16 + 120 = 137$ eigenvalues $n^{\zeta}_N(g)$ of the overlap matrix from the full set of projected 0qp, 2qp and 4qp configurations. Employing a logarithmic scale and ordering the eigenvalues increasingly, one observes that they gather in two distinct groups, i.e. one finds $17=n_{0{\rm qp}} + n_{2{\rm qp}}$ very small eigenvalues consistent with numerical noise and $120=n_{4{\rm qp}}$ values of order unity. Very naturally, the threshold is set such that only the latter eigenstates are kept to eventually diagonalize the Hamiltonian. Naively, the observation that the number of useful orthonormal states is strictly equal to the cardinal of projected 4qp states may suggest that the latter capture from the outset the information contained in the set of 0qp and 2qp configurations. Let us now investigate this hypothesis.

\begin{figure}[htbp] 
\includegraphics[width=\linewidth]{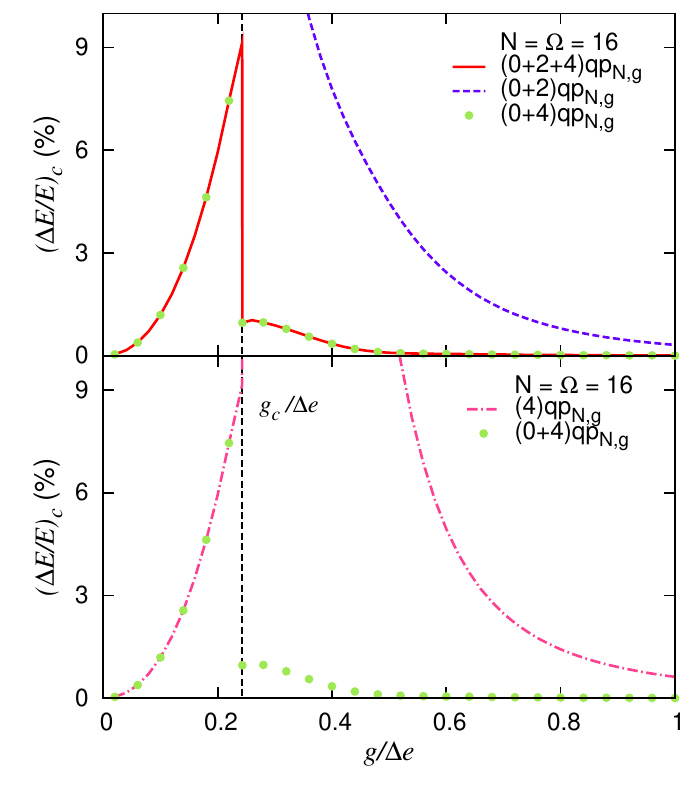} 
\caption{(color online) $(\Delta E / E)_c$ from truncated CI calculations as a function of $g$ for $N=\Omega=16$. Upper panel: results are shown for the full set of projected 0qp, 2qp and 4qp configurations (red solid line) as well as using 0qp and 4qp (green filled circles) or  0qp and 2qp (purple dashed line) configurations only. Lower panel:  results are shown for 0qp and 4qp configurations (green filled circles) as well as for 4qp configurations only (pink dashed-dotted line).}
\label{redundancyerror1} 
\end{figure}

Removing all 2qp configurations from the calculations, the middle panel of Fig.~\ref{eigenvaluesnorm} shows that only one small eigenvalue remains while $120=n_{4{\rm qp}}$ of them are still of order unity. Additionally, the upper panel of Fig.~\ref{redundancyerror1} testifies that the error on the correlation energy is the same as in the presence of projected 2qp configurations, which indeed appear to be redundant and can be entirely omitted from the outset. For large bases/particle number, the numerical scaling is governed by the number of 4qp configurations such that omitting projected 2qp excitations does not lead to a significant gain. Having one zero eigenvalue left, one may be tempted to conclude that the projected BCS reference state can be further removed from the linear combination. However, and as shown in the lower panel of Fig.~\ref{redundancyerror1}, the error on the correlation energy is huge for $g>g_c$ in this case. Thus, projected 4qp configurations do not fully contain the information built in the projected BCS state such that the useful set of $n_{\rm st} - 1 = n_{4{\rm qp}}$ orthonormal states do mix in a significant fraction of the projected 0qp state that cannot be plainly omitted. Ironically, bringing back projected 2qp configurations while keeping the projected BCS state aside is sufficient to gain  back the accuracy of the calculation based on projected 0qp and 4qp configurations, i.e. the set of projected 2qp configurations do bring in the mandatory information otherwise contained in the projected 0qp state. Of course, it is more efficient to do it by including 1 0qp state rather than 16 2qp configurations.

To eventually confirmed that all combinations of projected states are not equivalent, let us finally keep projected 0qp and 2qp configurations while omitting projected 4qp ones. In this case, one is left with $16=n_{2{\rm qp}}$ eigenvalues of order unity and a null one as shown in Fig.~\ref{eigenvaluesnorm}. As for the error on the correlation energy, the results are however much inferior to the full calculation as seen in the upper panel of  Fig.~\ref{redundancyerror1}. 

In conclusion, the information carried by projected 4qp states cannot be brought in by lower-order projected qp configurations while  the opposite is true to some extent.

\section{Optimized order parameter}
\label{Optimized}

As described in Sec.~\ref{optorderparam}, the order parameter of the BCS reference state associated with the underlying breaking of $U(1)$ symmetry can be optimized, for each "physical" $g$ of interest, when applying the truncated CI method. To do so, the diagonalization of $H(g)$ is repeated while scanning $g_{{\rm aux}}$ (i.e. $\Delta(g_{{\rm aux}})$) that parametrizes the truncated basis until the minimum of the lowest eigenenergy $E^{g_{{\rm opt}}}(g)$ is found. 

\subsection{PAV-BCS ansatz}

\begin{figure}[htbp] 
\includegraphics[width=\linewidth]{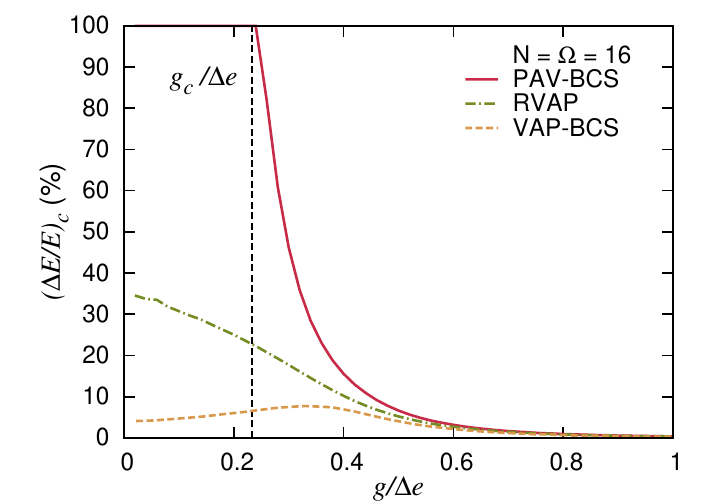} 
\caption{(color online) $(\Delta E / E)_c$ as a function of $g$ for $N=\Omega=16$. Results are shown for the the VAP-BCS (brown dashed line), the standard PAV-BCS (dark red solid line) and for a PAV-BCS calculation based on an optimized order parameter (dark green dashed-dotted line). The latter corresponds to the RVAP method.}
\label{fig:ervap} 
\end{figure}

As a jumpstart, the rationale is first applied while restricting the trial state to the first term in Eq.~\ref{eq:var1}, i.e. to the PAV-BCS wave-function. This strictly corresponds to the RVAP method designed within the frame of multi-reference nuclear energy density functional calculations~\cite{Rod05}. Results as a function of $g$ are compared in Fig.~\ref{fig:ervap} to actual PAV-BCS and VAP-BCS results. By definition, PAV-BCS results are generated by setting $g_{{\rm aux}}=g$ for each given $g$, i.e. by picking the order parameter obtained at the level of the BCS wave-function rather than at the level of the actual PAV-BCS wave-function.

While the results are not at the desired level because of the lack of projected qp excitations, they perfectly illustrate the gain induced by optimizing the order parameter at the level of the full calculation, i.e. after the symmetry restoration is performed in the present example rather than prior to it. It is particularly striking below threshold where  $(\Delta E / E)_c$ decreases from 100 \% to about 20-40 \%. In the normal phase, not too far from the BCS threshold, it is indeed highly beneficial to allow the reference state to break $U(1)$ symmetry while restoring it. As discussed above, this corresponds to including a specific set of np-nh configurations at a low computational cost. This reduced set provides an efficient way to partly capture correlations associated with pairing fluctuations that arise as a precursor of the phase transition. Above threshold, results are also significantly improved over the range $g \in [g_c,0.4]$ by finding the optimal order parameter. For $g>0.6$, no significant gain is obtained given that PAV-BCS itself becomes eventually exact.

One interest of this optimization is that the associated numerical effort simply corresponds to repeating the full calculation a few number of times. At the PAV level, it makes the RVAP calculation unexpensive compared to the VAP-BCS calculation it approximates. Of course, results are significantly less accurate than the actual VAP-BCS calculation given that the optimization of the order parameter is not equivalent to exploring the complete manifold of BCS states as in the VAP-BCS calculation. This is particularly true in the very weak coupling regime where the system does not experience pairing fluctuations.

\subsection{Full ansatz}

\begin{figure}[htbp] 
\includegraphics[width=0.9\linewidth]{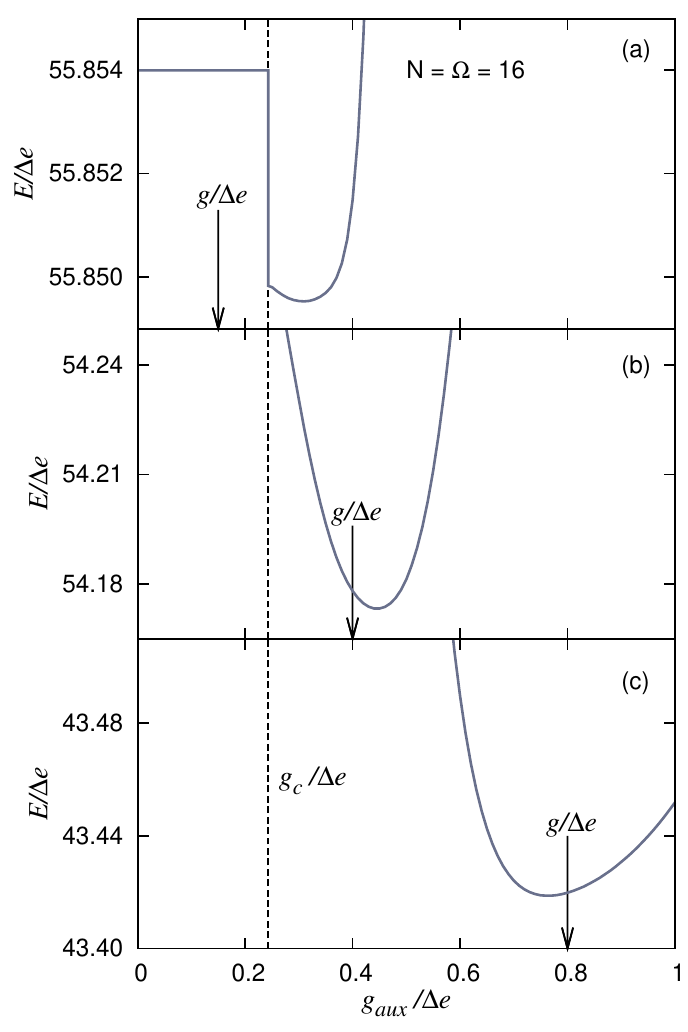} 
\caption{(color online) Total binding energy from truncated CI calculations based on projected 0qp, 2qp and 4qp configurations as a function of $g_{{\rm aux}}$. The calculations are performed for $N=\Omega=16$ and the specific value of $g$ indicated by the arrow. Panel (a): $g=0.15 < g_c$. Panel (b): $g=0.4 > g_c$. Panel (c): $g=0.8 \gg g_c$.
%The dashed line corresponds to $g_c = 0.24$.
}
\label{fig6:juju} 
\end{figure}

The rationale is now implemented on the basis of the full ansatz of Eq.~\ref{eq:var1}. Figure~\ref{fig6:juju}  displays the so-called potential energy surface (PES) representing the total energy  $E^{g_{{\rm aux}}}(g)$ as a function of $g_{{\rm aux}}$. Results are given for three representative values of $g$, i.e. (a) $g=0.15 < g_c$, (b) $g=0.4 > g_c$ and (c) $g=0.8 \gg g_c$. 

In each case, the minimum of the PES indicates the position of  $g_{{\rm opt}}$. One first notices that the minimum of the curve is typically not obtained for $g_{{\rm aux}}=g$. The optimal basis in presence of the configuration mixing is characterized by a symmetry breaking, i.e. a reference pairing gap $\Delta(g_{{\rm opt}})$, that differs from the one obtained at the (projected) BCS minimum. This is particularly striking for $g<g_c$ (upper panel of Fig.~\ref{fig6:juju}) where it is advantageous to employ a basis that explicitly captures features of pairing fluctuations, i.e. that benefits from the additional np-nh configurations brought about by projected 0qp, 2qp and 4qp states. Beyond the phase transition, one has $g_{{\rm opt}}>g$ ($g_{{\rm opt}}<g$) for intermediate (large) coupling as exemplified in the middle (lower) panel of Fig.~\ref{fig6:juju}. All in all, the successive inclusion of the particle number restoration and of the qp excitations significantly influence the value of $g_{{\rm opt}}$ and the associated quality of the variational ansatz (see below), especially at weak and intermediate coupling. This is summarized in Tab.~\ref{variousgopt}.

\begin{table}[htbp]
  \centering 
  \begin{tabular}{|c|c|c|c|}
\hline
$g_{{\rm opt}}$ & BCS & PAV-BCS & Truncated CI \\
\hline
$g=0.15$ & $g$ & 0.29 & 0.31 \\
$g=0.4$  & $g$ & 0.44 & 0.45 \\
$g=0.8$  & $g$ & 0.82 & 0.76 \\
\hline
\end{tabular}
  \caption{Optimal order parameter $g_{{\rm opt}}$ of the reference state at various level of approximation, i.e. BCS, PAV-BCS and truncated CI calculation based on projected 0qp, 2qp and 4qp configurations, for $N=16$. Results are provided for $g=0.15 < g_c$, $g=0.4 > g_c$ and $g=0.8 \gg g_c$. We recall that $g_c=0.24$ for $N=16$.}
  \label{variousgopt}
\end{table}

\begin{figure}[htbp] 
\includegraphics[width=\linewidth]{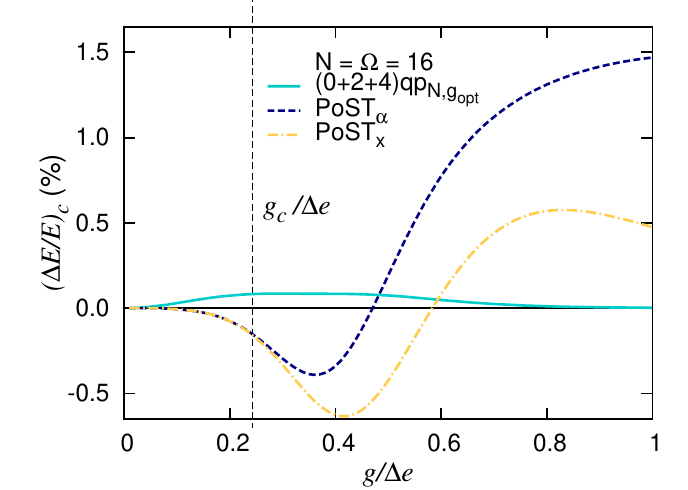} 
\caption{(color online) Same as Fig.~\ref {fig2:juju} but with the optimal order parameter $g_{{\rm opt}}$ defining the basis at each value of the coupling $g$ for the truncated CI (light blue solid line). Results of PoST$_\alpha$ (dark blue dashed line) and PoSTx~\cite{Deg16} (yellow dashed-dotted line) are shown for comparison.}
\label{fig7:juju} 
\end{figure}

Figure~\ref{fig7:juju} provides the same comparison as Fig.~\ref {fig2:juju} but with the optimal order parameter $g_{{\rm opt}}$ defining the basis at each value of the coupling $g$. The optimization generates an impressive systematic improvement for $g<0.6$ and solves completely the discontinuity problem observed in Fig. \ref{fig7:juju} at $g=g_c$. The error on the correlation energy is now lower than 0.1 \% for all $g$, which is almost one order of magnitude lower than our original goal. Our results compare very favorably with PoST methods~\cite{Duk16,Deg16}. Once again, projected 2qp configurations are redundant and can actually be omitted.

\begin{figure}[htbp] 
\includegraphics[width=0.9\linewidth]{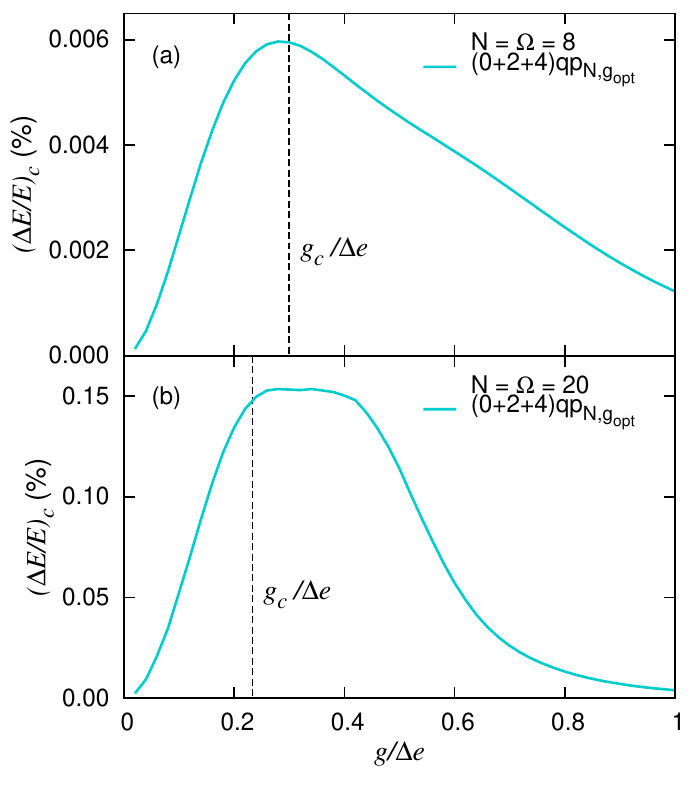} 
\caption{(color online) Same as Fig. \ref {fig7:juju} for (a) $N=\Omega=8$ and (b) $N=\Omega=20$.}
\label{fig8:juju} 
\end{figure}

Figure.~\ref{fig8:juju} displays similar results for (a) $N=\Omega=8$ and (b) $N=\Omega=20$. Conclusions are essentially the same as for $N=\Omega=16$. 

\section{Performance and scaling}
\label{Comparison}

The main feature of the presently proposed method resides in the optimization of the basis used to diagonalize the Hamiltonian. This results in a dimensionality that is drastically reduced compared to the total Hilbert space and, for a given accuracy, compared to truncated CI calculations based on traditional np-nh configurations. The rationale of the latter method is to describe the system via a basis of product states that respect $U(1)$ symmetry even in the superfluid phase. The rationale of our method is exactly opposite, i.e. it uses a basis that exploits the breaking of $U(1)$ symmetry (while restoring it) to describe the system even in its normal phase.  

Table~\ref{tab:hilbert} compares, for $N=16$, the total size of ${\cal H}_N$ to the cardinal of the basis employed in standard np-nh truncated CI calculations up to 8p-8h~\cite{Pil05}, as well as in the presently designed approach. The corresponding error on the correlation energy is provided for $g=0.18$, $g=0.54$ and $g=0.66$. We recall in passing that the dimension of the basis used in truncated CI calculations based on 0qp and 4qp configuration makes the method exact for $N=2$ and $N=4$.

In the weak coupling regime ($g=0.18$), truncated CI calculations based on 0p-0h and 2p-2h configurations already achieve an error below 1 \% based on a  small basis size, which eventually scales as $N^2$ with the system size. Calculations based on optimized projected 0qp and 4qp configurations perform one order of magnitude better based on a basis that is only twice as large and that scales similarly with the system size. If degrading the calculation to optimized projected 0qp and 2qp configurations, a scheme that scales as $N$ with the system size, the result are however one order of magnitude worse (10 \% error on $E_c$) than the CI calculation based on 0p-0h and 2p-2h configurations. This demonstrates the need to include 4qp configurations to reach (much) better than the 1 \% accuracy at weak coupling.

In the superfluid regime, truncated CI calculations based on projected 0qp and 4qp configurations reach again an accuracy well below 1 \%, which is comparable to the results obtained from truncated CI calculations including up to 8p-8h configurations for $g=0.54$ and is even one order of magnitude better for $g=0.66$. While the dimension of the latter basis scales as $N^8$ with the system size, the set of projected 0qp and 4qp configurations scales as $N^2$, which is obviously much more gentle. For rather strongly paired systems, i.e. for $g=0.66$, degrading the calculation to optimized 0qp and 2qp configurations, which scales as $N$ with the system size, already reaches 1 \% accuracy on the correlation energy.

\begin{table}[htbp]
  \centering 
  \begin{tabular}{| c || c | c| c| c|}
\hline
$N=16$ &   $n_{{\rm st}}$ & $g=0.18$ & $g=0.54$ & $g=0.66$ \\
\hline
\hline
 $2$p-$2$h & 65   & 0.64 $\%$  & 20.92  $\%$     & 29.37 $\%$\\
 $4$p-$4$h & 849  & 0.01 $\%$  & 5.22 $\%$      & 9.59  $\%$\\
 $6$p-$6$h & 3985 & 0.00 $\%$  &  0.60 $\%$  & 1.66  $\%$\\
 $8$p-$8$h & 8885 & 0.00 $\%$  & 0.03 $\%$  & 0.12  $\%$\\
\hline
 $(0\!+\!2){\rm qp}_{N,g}$   & 17 & 100  $\%$  & 3.52 $\%$ &  1.70 $\%$ \\
 $(0\!+\!4){\rm qp}_{N,g}$   & 121 &  0.64 $\%$ &  0.07 $\%$  &  0.04 $\%$ \\
 $(0\!+\!2\!+\!4){\rm qp}_{N,g}$   & 137 & 0.64  $\%$ & 0.07 $\%$  & 0.04  $\%$ \\
\hline
 $(0\!+\!2){\rm qp}_{N,g_{{\rm opt}}}$   & 17 &  9.20 $\%$  & 3.34 $\%$ &  1.66 $\%$ \\
 $(0\!+\!4){\rm qp}_{N,g_{{\rm opt}}}$   & 121 &  0.07 $\%$ & 0.07  $\%$  & 0.03  $\%$ \\
 $(0\!+\!2\!+\!4){\rm qp}_{N,g_{{\rm opt}}}$   & 137 &  0.07 $\%$ & 0.07 $\%$  & 0.03  $\%$ \\
\hline
 Exact     & 12870 & 0.00   $\%$ & 0.00 $\%$ & 0.00 $\%$\\
\hline
\end{tabular}
  \caption{Dimensionality $n_{{\rm st}}$ of the full $N$-body Hilbert space  ${\cal H}_N$ for $N=\Omega=16$ as well as of the sub-space considered in np-nh truncated CI calculations as well as in our method. In each case, the last three columns display the error on the correlation energy for $g=0.18$, $g=0.54$ and $g=0.66$, respectively.}
  \label{tab:hilbert}
\end{table} 

Of course, part of the cost of the calculation is transferred into the particle-number projection but the end scaling is still very favorable. Eventually, the numerical cost ${\rm Num}(N)$ of the scheme is polynomial and scales according to 
\begin{eqnarray}
{\rm Num}(N) &=& n_{g_{{\rm aux}}}\Big( {\rm BCS}(g_{{\rm aux}},N) \nonumber \\
&& \hspace{1cm} +  n^2_{\rm st} \, {\rm ME}(n_{\phi},N) \nonumber \\
&& \hspace{1cm} +  {\rm DIAG}(n_{\rm st}) \Big) \, , 
\end{eqnarray}
where the first term relates to solving BCS equations, the second term to calculating the elements of the overlap and Hamilton matrices while the third term designates the cost of the diagonalization of these two matrices.  

The cost scales linearly with the number of times $n_{g_{{\rm aux}}}$ the calculation must be performed to find the optimal $g_{{\rm opt}}$. In practical calculations, it is possible to keep $n_{g_{{\rm aux}}}<10$ once the calculation at $g_{{\rm aux}}=g$ has been performed. Of course, $n_{g_{{\rm aux}}}=1$ when the optimization of the order parameter characterizing the basis is omitted. 

The cost associated with the BCS variation is negligible as it scales essentially linearly with $\Omega=N$. Employing projected 0qp and 4qp (2qp) configurations, the number of matrix elements $n^2_{\rm st}$ to calculate scales as $N^4$ ($N^2$) while the cost of their computation is ${\rm ME}(n_{\phi},N) = \alpha \, n_{\phi}\,N^2$, which makes the overall scaling go as $n_{\phi}\,N^6$ ($n_{\phi}\,N^4$). The cost of computing the matrix elements is linear with the number of gauge angles $n_{\phi}$ employed in the particle-number projector (see Eq.~\ref{projector}). This number can be kept essentially constant, i.e. $n_{\phi}\sim 10$, when increasing $N$. Finally, the cost of the diagonalization is ${\rm DIAG}(n_{\rm st})= \beta n^3_{\rm st} = \beta  N^6$ ($\beta N^3$). 

All in all, the building of the matrices and their diagonalization scale similarly as $N^6$ (the building of the matrix goes as $N^4$ and dominates when using projected 0qp and 2qp configurations) with the system size. There are ways to further improve on this situation. First, full diagonalization is not mandatory as one can envision the use of alternative methods such as Lanczos to extract a few low-lying states at a much reduced numerical cost. This might be particularly useful when addressing large model spaces and/or particle numbers associated with realistic cases of interest. Second, the pre-factor $\alpha\,n_{\phi}$ associated with the direct integration over the gauge angle to perform the particle number projection can be scaled down by performing the latter on the basis of recurrence relations~\cite{Hup11a}. 

Last but not least, there probably is a systematic convergence of the result, as in standard truncated CI calculations~\cite{Pil05}, as a function of the maximum unperturbed energy of the 2qp and 4qp included in the ansatz for a given single-particle basis size ($\Omega$ here). This means that given a targeted accuracy, the dimensionality and the numerical cost might be significantly scaled down by exploiting this additional convergence parameter and complementing the calculation by an appropriately designed formula to extrapolate the results to the un-truncated limit. Such a systematic study has not been performed within the scope of the present paper but could be envisioned in the future.

\section{Additional observables}
\label{Observables}

To complete our study, the discussion is extended to other observables. 

\subsection{Effective pairing gap}

We start with the computation of the effective pairing gap~\cite{vondelft96a,braun98a}
\begin{eqnarray}
\Delta_{\rm  eff}(g) & = & g \sum_{k=1}^{\Omega} \sqrt{\langle a^\dagger_k a^\dagger_{\bar k} a_{\bar k}  a_k \rangle %\nonumber \\
%&-& 
-\frac{1}{4} \langle(a^\dagger_k a_k + a^\dagger_{\bar k} a_{\bar k} ) \rangle^2 } \, , \label{eq:effgap}
\end{eqnarray}
which generalizes the BCS gap $\Delta(g)$ and where the expectation values are to be computed for any ground-state wave-function of interest.

\begin{figure}[htbp] 
\includegraphics[width=0.9\linewidth]{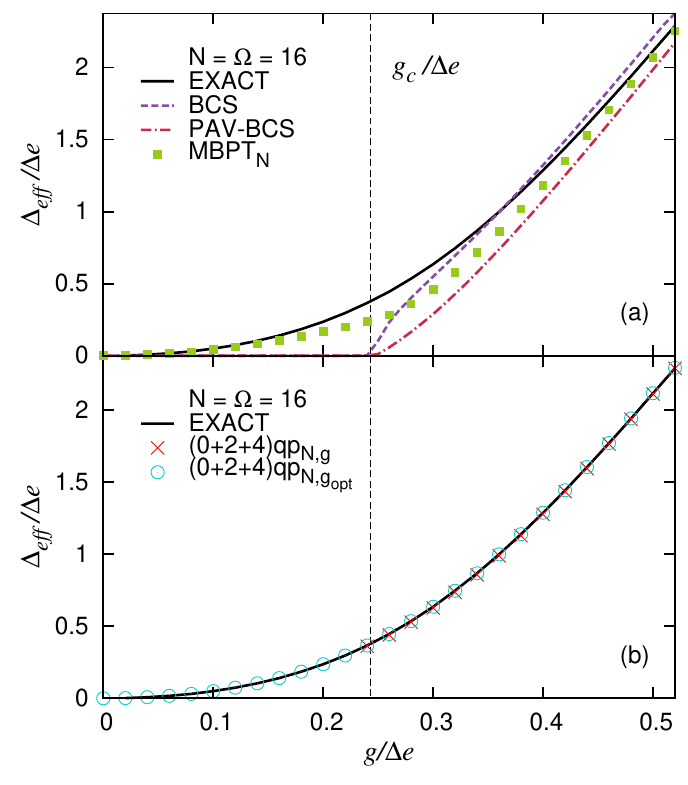} 
\caption{(color online) Ground-state effective pairing gap (Eq.~\ref{eq:effgap}) as a function of $g$ for $N=16$. Top panel: exact results (black solid line) against BCS (purple dashed line), PAV-BCS (red dot-dashed line) and MBPT$_{N}$ (green filled squares). Lower panel: exact results against truncated CI based on non-optimized (red cross) or optimized (blue circles) projected 0qp, 2qp and 4qp configurations. }
\label{fig:effgap} 
\end{figure}

In Fig \ref{fig:effgap}, the effective gap obtained in the exact case is compared to the one obtained from various approximate many-body methods of present interest. We observe that truncated CI calculations based on (non) optimized projected 0qp and 4qp configurations provide results that are below 0.05 \% (1.5 \%) error for all coupling strengths $g$ ($g>g_c$) and much superior to the other methods shown.
 
\subsection{One-body entropy}

\begin{figure}[htbp] 
\includegraphics[width=0.9\linewidth]{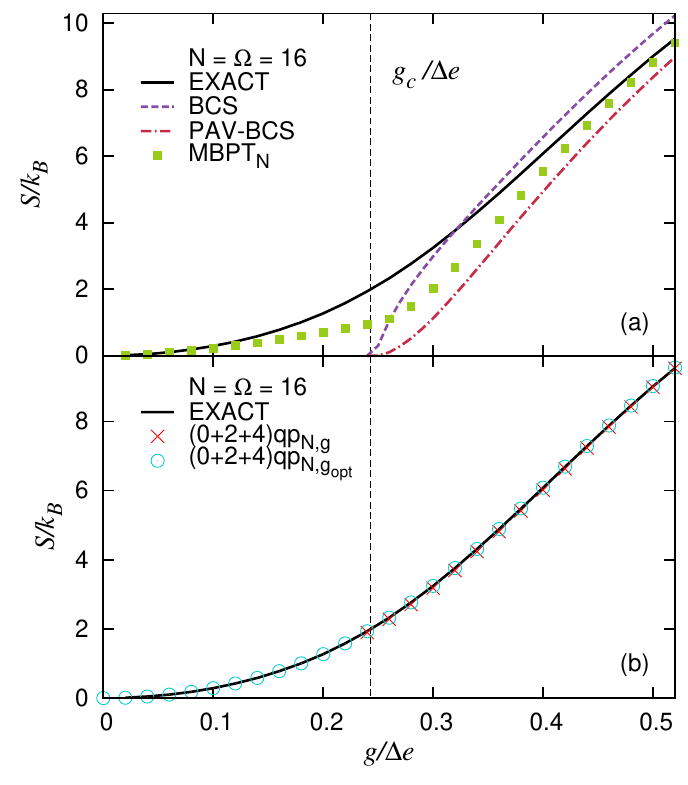} 
\caption{(color online) Same as Fig.~\ref{fig:effgap} for the one-body entropy.}
\label{fig:entropy} 
\end{figure}

States obtained via the presently proposed method are strongly entangled, in the sense that they correspond to a complex mixing of independent-particle states. As a matter of fact, exact solutions are known to be highly correlated states, resulting into extended  diffusion of single-particle occupation numbers across
the Fermi energy. To quantify the deviation of these many-body states from any independent-particle state, the single-particle entropy defined as
\begin{eqnarray}
\frac{S}{k_B} & = & \!- 2  \sum_{k=1}^{\Omega} \Big\{ \langle a^\dagger_k a_k  \rangle \ln  \langle a^\dagger_k a_k  \rangle
                         \nonumber \\
                         && \hspace*{1.cm} \!+\! (1\!-\!\langle a^\dagger_k a_k  \rangle) \ln (1\!-\!\langle a^\dagger_k a_k  \rangle) \Big\}\, .
%S & = & - 2 k_B \sum_{k=1}^{\Omega}  \left\{ n_i \ln n_i + (1-n_i) \ln (1-n_i) \right\}
\label{eq:spentrop}
\end{eqnarray}  
is computed. Exact results are compared in Fig.~\ref{fig:entropy}  to those obtained from various approximate many-body methods of present interest. Again, truncated CI calculations based on (non) optimized projected 0qp and 4qp configurations provide results that are below 0.1 \% (2 \%) error for all coupling strengths $g$ ($g>g_c$) and much superior to the other methods shown. This demonstrates that single-particle occupation numbers across the Fermi energy are accurately described, which eventually propagate to any one-body observable.  
 
\subsection{Low-lying excitations}
\label{Spetro}

Only ground-state properties have been discussed so far. Being based on a direct diagonalization of the Hamiltonian in a restricted space, it is a tremendous advantage of the presently designed method to also access excited states. Given that the size of the sub-space covered is drastically smaller than the one of ${\cal H}_N$, one can only expect to provide a fair account of a few low-lying states. 

\begin{figure}[htbp] 
\includegraphics[width=0.9\linewidth]{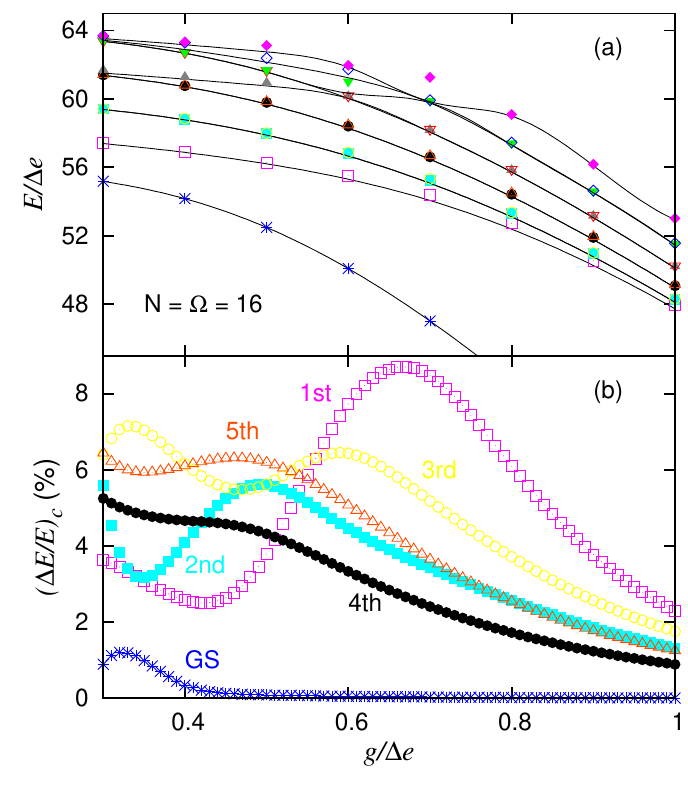} 
\caption{(color online) Excitation energy as a function of $g$ of low-lying excited states obtained for $N=\Omega=16$ with $g_{{\rm aux}}=g$. Exact results (black solid lines) are compared to truncated CI calculations based on projected 0qp, 2qp and 4qp configurations (symbols). Ground state results are also shown for comparison. Upper panel: absolute energies of the 10 lowest excited states. Lower panel: relative errors on the correlation energy of the 5 lowest excited states.}
\label{fig11:juju} 
\end{figure}  
   
Energies of the 10 lowest excited seniority-zero states are compared to exact results for $N=\Omega=16$ in Fig.~\ref{fig11:juju}. Truncated CI calculations based on projected 0qp, 2qp  and 4qp configurations provide an accurate reproduction of the low-lying spectroscopy. More specifically, the error made on the correlation energy\footnote{The correlation energy of a given excited state is defined by subtracting the energy of the HF (excited) configuration obtained from the associated exact Richardson solution as $g$ goes to 0.} of the five lowest excited states is lower than 8.5 \% for $g \in [0,1]$. This error drops to less than 3.5 \% for $N=8$. 

It is to be remarked that the order parameter has not been optimized, i.e. $g_{{\rm aux}}=g$ is presently used, because such an optimization does not necessarily decrease the error. Indeed, the improvement of the ground-state energy obtained on the basis of  Ritz' variational principle does not carry over to excited states. While no particular pattern can be anticipated for individual excited states, it happens that the reproduction of the low-lying spectroscopy is of the same overall quality for $g_{{\rm aux}}=g$ and $g_{{\rm aux}}=g_{{\rm opt}}$.

\section{Conclusions}
\label{Conclu}

A novel approximate many-body scheme is presently tested on the so-called attractive pairing Hamiltonian as a way to gauge its capacity to account for the physics of $N$-body systems transitioning from the weak to the strong coupling regime via a normal-to-superfluid phase transition. This work takes place in the context of designing polynomially-scaling methods that are possibly more (i)  accurate and (ii) easily applicable to more quantum states than those, i.e. Gorkov self-consistent Green's function (GSCGF), multi-reference in-medium similarity renormalization group (MR-IMSRG) and Bogoliubov coupled cluster (BCC) methods, that are currently operating a breakthrough in the  ab-initio calculations of medium-mass open-shell nuclei. 

The presently proposed method is variational and happens to be an interesting candidate to achieve the above-mentioned goal. It does so by combining three features that have been employed separately in various existing many-body methods so far
\begin{itemize}
\item It is a truncated configuration interaction method, i.e. it amounts to diagonalizing the Hamiltonian in a highly truncated subspace of the total $N$-body Hilbert space. 
\item The reduced Hilbert space is generated via a set of states that exploit the spontaneous ($U(1)$) symmetry breaking and restoration associated with the (normal-to-superfluid) quantum phase transition of the $N$-body system. Specifically, the set of states considered is given by the particle-number projected BCS state along with projected seniority-zero two and four quasi-particle excitations built on top of the BCS state. Because each basis state is symmetry projected, the method consists of representing the Schroedinger equation onto a {\it non-orthonormal} basis. The corresponding diagonalization can be performed using standard techniques. 
\item The extent by which the BCS reference state breaks ($U(1)$) symmetry is optimized {\it in presence} of projected two and four quasi-particle excitations. This constitutes an extension of the so-called restricted variation after projection method in use within the frame of multi-reference nuclear energy density functional calculations~\cite{Rod05}.
\end{itemize}

The many-body scheme has been compared to exact solutions of the attractive pairing Hamiltonian based on Richardson equations~\cite{Ric64,Duk04,Van06}. By construction, the method is exact for $N=2$ and $N=4$. For $N=(8,16,20)$, the error on the ground-state {\it correlation energy} is less than (0.006, 0.1, 0.15) \% across the entire range of coupling $g$ defining the pairing Hamiltonian and driving the normal-to-superfluid quantum phase transition. To the best of our knowledge, this is better than any many-body method scaling polynomially ($N^6$ here) with the system size and tested so far on the pairing Hamiltonian. In particular, it is superior to the highly accurate PoST$_\alpha$ and PoST$_x$ methods recently proposed in Ref.~\cite{Deg16} with the same motivations as here. The presently proposed method offers the great additional advantage to automatically access low-lying excited states. The error on the correlation energy of the five lowest excited states is smaller than 8.5 \% (3.5 \%) for $g \in [0,1]$ for $N=16$ ($N=8$).

The schematic pairing Hamiltonian employed here corresponds to modeling sub closed-shell systems, i.e. the naive filling of the doubly-degenerate picket fence single-particle scheme with an even number of particles always leads to a sub closed-shell system. Correspondingly, the Hartree-Fock reference state can always be defined, which is mandatory to apply many methods, including the recently proposed PoST methods~\cite{Deg16}. However, this HF reference cannot even be defined in genuinely open-shell systems we are actually interested in, i.e. for the vast majority of singly or doubly open-shell nuclei. The presently proposed method, however, is based on a reference state that spontaneously breaks ($U(1)$) symmetry whenever necessary and can be equally applied independently of the closed-shell, sub closed-shell or genuinely open-shell character of the system under study. This makes the method extremely versatile.

Although IMSRG and SCGF techniques have not been applied to the pairing Hamiltonian problem throughout the superfluid phase transition (while CC has), their accuracy in the best current level of implementation is of the order of a few per cent error on the ground-state correlation energy of singly open-shell nuclei. In view of that, results obtained in the present work indicate that the truncated CI method based on low-order projected qp excitations constitutes an interesting method to pursue. In order to go beyond the present proof-of-principle calculation, our objective is to implement the method for ab initio calculations of mid-mass open-shell nuclei.

Last but not least, one should note that the highly accurate character of the method is achieved at the price of giving up on size-extensivity. It is a common feature of all truncated CI methods that is also shared by the PoST method of Ref.~\cite{Deg16}. The increasing relative error from 0.006\%, to 0.1 \% and to  0.15 \% when increasing the particle number from $N=8$ to $N=16$ and to $N=20$ might already be a trace of it. Restoring size consistency demands the inclusion of very high excitation levels and possibly all excitations, which is prohibitive. Although given up on size extensivity is somewhat unconventional from the perspective of modern many-body methods, and although it deserves attention as larger systems are studied, it is a price one is willing to pay to obtain a highly accurate description at a reasonable computational cost.
\newline

\section*{Acknowledgement}
\label{acknow}

The authors thank M. Degroote, T. M Henderson, J. Zhao, J. Dukelsky and G. E. Scuseria for useful discussions and for providing their data from the PoST methods. The authors thank B. Bally for proofreading the manuscript.

\bibliography{paper_Julien}

\end{document}